\documentclass[12pt]{article}
\usepackage{epsf}
\newcommand{\R}{\rm I\kern-.2emR}
\newcommand{\C}{\rm \kern.25em\vrule height1.4ex
 depth-.12ex width.06em\kern-.31em C}
\newcommand{\N}{{\rm I\kern-.16em N}}
\newcommand{\Z}{{\rm Z\kern-.35em Z}}
\newcommand{\bee}{\begin{equation}}
\newcommand{\ee}{\end{equation}}
\newcommand{\ba}{\begin{array}}
\newcommand{\ea}{\end{array}}
\newcommand{\bea}{\begin{eqnarray}}
\newcommand{\eea}{\end{eqnarray}}

\begin{document}
\begin{flushright}                                                              
MPI-PhT/98-77 \\
\end{flushright}                                                                
\bigskip\bigskip\begin{center}
{\Huge Critical behavior of classical spin models and local cohomology
\footnote{Talk given at the XIIth Max Born Symposium, Wroclaw, dedicated
to Jan Lopuszanski on the occasion of his 75th birthday.}
}
\end{center}  
\vskip 1.0truecm
\centerline{Adrian Patrascioiu}
\centerline{\it Physics Department, University of Arizona}
\centerline{\it Tucson, AZ 85721, U.S.A.}
\vskip5mm
\centerline{and}
\vskip5mm
\centerline{Erhard Seiler}
\centerline{\it Max-Planck-Institut f\"{u}r Physik}
\centerline{\it  (Werner-Heisenberg-Institut) }
\centerline{\it F\"ohringer Ring 6, 80805 Munich, Germany}
\bigskip \nopagebreak \begin{abstract}
\noindent
Using reflection positivity as the main tool, we establish a
connection between the existence of a critical point in classical spin 
models and the triviality of a certain local cohomology class related to 
the Noether current of the model in the continuum limit. Furthermore we
find a relation between the location of the critical point and the
momentum space autocorrelation function of the Noether current.
\end{abstract}
\vskip 5mm
\newpage
\vskip4mm \noindent
{\bf 1. Introduction: Lattice and Continuum}
\vskip2mm
It is well known that one possible approach to the construction of a
Quantum Field Theory (QFT) goes by way of taking the continuum limit of a
system of Classical Statistical Mechanics on a lattice, such as the
Ising model, the classical Heisenberg model or more generally a
classical spin model. Taking the continuum limit means in this context
that one has to drive the system to a critical point, that is a point
at which the dynamically produced scale(s) become infinite in terms
of the lattice spacing; the continuum limit is then obtained by
an infinite rescaling of the lattice model (see below; a rather detailed 
discussion of how this is done is contained in \cite{PScont}). A bonus of
this construction is that the continuum limit inherits certain properties
of the lattice model, such as Reflection Positivity (RP) which leads to 
positivity of the state space metric and the spectrum condition of the 
QFT.

More precisely we have to distinguish between two kinds of continuum
limits:
\begin{itemize}
\item The massive continuum limit: one chooses the dynamically
generated length (correlation length) $\xi$ of the system as the standard
of length, considers the system at length scales that are fixed
multiples of that standard, and sends $\xi\to\infty$ by driving the 
system to criticality.
\item The massless continuum limit: the lattice system is put right on
a critical point; one then chooses an arbitrary length scale that
becomes infinite in lattice units and rescales the system accordingly.
\end{itemize}

The first option will produce a (Euclidean) QFT with unit mass, the
second one a massless QFT, which according to standard lore will also
be conformally invariant. In $2D$ it is believed that this allows to
classify the critical behaviors according to the well-studied
(rational) Conformal QFTs (\cite{BPZ,FQS}).

In this talk we want to discuss this connection, and actually close some 
gaps. In the course of the argument it turns out that one has to prove 
the triviality of a certain `local cohomology class' related to the 
Noether current. This is possible with the use of lattice Ward identities
(WI) and RP.

The same ingredients lead at the same time to an interesting and maybe
unexpected relation between the location of the critical point of the
lattice model and the Noether current 2-point function of the continuum
model. This leads to a new criterion that allows to discriminate between
the `conventional wisdom' about nonabelian spin models in $2D$, which
posits that they do not become critical at any temperature and the 
scenario long advocated by us \cite{ps} that they do have a transition 
to a massless spin wave phase, just as the plane rotator model.
\newpage
\vskip4mm\noindent
{\bf 2. Local Cohomology}
\vskip2mm

It has been noted long ago \cite{strocchi,pohlmeyer,roberts} that the 
imposition of locality (local commutativity, Einstain causality)
may make the cohomology of Minkowski space nontrivial.

The problem of local cohomolgy may be stated as follows: assume that an
antisymmetric tensor field $\Phi_{\mu_1,...,\mu_k}(x)$ is given,
which satisfies Wightman's axioms and is closed, i.e. satisfies
\bee
d\Phi\equiv d(\sum \Phi_{\mu_1,...,\mu_k} dx^{\mu_1}...dx^{\mu_k})=0
\ee
(in the notation of alternating differential forms).

The question is then under which conditions the field $\Phi$ is
exact, i.e. there exists a local antisymmetric tensor field $\Psi$
such that $\Phi=d\Psi$.

There are some well-known examples where the answer is `no', even though
Minkowski space is topologically trivial:

(1) the free Maxwell field $F$ in dimension $D\ge 2$ \cite{strocchi};

(2) the gradient of the massless free scalar field $\phi$ in $2D$, 
because the field $\phi$ does not exist as a local (Wightman) field.

There is also a simple $2D$ example on which we hit in our analysis
of $2D$ classical spin models: let
\bee
\Phi=\phi_c dx^1dx^2
\ee
where $\phi_c$ has the Euclidean two-point function
\bee
\langle \phi_c(0)\phi_c(x)\rangle={1\over (x^2)^2}.
\ee
Then $\Phi$ is trivially closed in $2D$, but it is not exact, i.e. there
is no local vector field $j_\mu$ such that
\bee
\phi_c=\epsilon_{\mu\nu}\partial_\mu j_\nu
\ee

This example can be made more explicit by requiring $\phi_c$ to be
a generalized free, i.e. Gaussian field, with its two-point function
given by eq.(3). If we solve the
differential equations that the two-point function of $j_\mu$ has to
fulfill in order to satisfy eq.(4) and impose euclidean covariance,
we find that there is no scale invariant solution. The covariant
solutions are
\bee
G_{\mu\nu}(x)=-\delta_{\mu\nu}{\ln x^2+\lambda \over 8x^2}
+ x_\mu x_\nu {\ln x^2+1+\lambda\over 4x^2}
\ee
This is not the two point function of a local vector field, continued
to euclidean times: it violates the so-called reflection positivity
\cite{OS}, because the logarithm changes sign. Similarly it also does
not obey the positivity required for a random field.

\vskip4mm\noindent
{\bf 3. What is the massles continuum limit of a critical classical
spin model?}
\vskip2mm

There is an old argument \cite{affleck} that a classical spin model with 
a continuous symmetry group $G$ will have a massless contiuuum limit that
has an enhanced $G\times G$ symmetry; this is supposed to come about
due to the splitting of the model into two independent `chiral' theories.
Affleck \cite{affleck} gave this argument in the the framework of 
Quantum Field Theory in Minkowski space, but it can be easily rephrased 
in the euclidean setting. In \cite{pso21, pso22} we pointed out two 
possible gaps in those arguments coming from hidden assumptions whose
validity ahs to be checked. But in those papers we also showed that these
gaps can be closed, using properties like reflection positivity.

The core of the euclidean version of Affleck's argument is the following:
assume that we have a scale invariant continuum theory with a conserved
current $j_\mu(x)$. Euclidean covariance requires that the two-point
function $G_{\mu\nu}$ of $j_\mu$ is of the form
\bee
G_{\mu\nu}\equiv\langle j_\mu(0) j_\nu(x)\rangle=\delta_{\mu\nu}
{b\over x^2}+{ax_\mu x_\nu\over(x^2)^2} \quad \ (x\neq 0)
\ee
Imposing current conservation means
\bee
\partial_\mu G_{\mu\nu}=0
\ee
for $x\neq 0$, which implies
\bee
a=-2b
\ee
\bee
G_{\mu\nu}(x)=b({\delta_{\mu\nu}\over x^2}
-{2x_\mu x_\nu\over (x^2)^2})
\ee
This is, up to the factor $b$, equal to the two point function of
$\partial_\mu\phi$ where $\phi$ is the massless free scalar field
(it is irrelevant here that the massless scalar field does not exist
as a Wightman field). If we look at the two-point function of the dual 
current $\epsilon_{\mu\nu}j_\nu$, it turns out to be
\bee
\tilde G_{\mu\nu}\equiv \epsilon_{\mu\lambda}\epsilon_{\rho\nu}
G_{\lambda\rho}=-G_{\mu\nu}
\ee
so the dual current two point function satisfies automatically the
conservation law. Conservation of the two currents $j$ and  $\tilde j$ is 
equivalent to conservation of the two chiral currents $j_\pm=j_0\pm j_1$
in Minkowski space.

By general properties of local quantum field theory (Reeh-Schlieder 
theorem, see \cite{RS,SW}) it follows that the dual current  is conserved as
a quantum field. So the two conservation laws together imply that
\bee
j_\mu=\sqrt{b} \partial_\mu\phi  ,
\ee
where $\phi$ is the massless scalar free field, and also that
\bee
j_\mu=\sqrt{b} \epsilon_{\mu\nu}\partial_\nu\psi  ,
\ee
where $\psi$ is another `copy' of the massless scalar free field.

As presented, this argument is certainly correct. But it depends on the
assumption that the Noether currents {\it exist} as Wightman fields,
and this assumption is in fact nontrivial and could a priori fail
in the critical spin models. A simple example of a Quantum Field
Theory with a continuous symmetry in which the Noether current does
not exist as a Wightman field is given by the two-component free field
in $2D$ in the massless limit. It is simply given by a pair of 
independent Gaussian fields $\Phi^{(1)},\Phi^{(2)}$, both with covariance
\bee
C(x)={1\over (2\pi)^2}\int d^2p {e^{ipx}\over p^2+m^2}.
\ee
where we are interested in the limit $m\to 0$.
This system has a global $O(2)$ invariance rotating the two fields into 
each other. It is well known that the massless limit only makes sense for 
functions of the gradients of the fields. But the Noether current of the 
$O(2)$ symmetry is given by
\bee
j_\mu(x)=\Phi^{(1)}(x)\partial_\mu\Phi^{(2)}(x)
-\Phi^{(2)}(x)\partial_\mu\Phi^{(1)}(x),
\ee
and it cannot be written as a function of the gradients. It is also easy 
to see directly that its correlation functions do not have a limit as 
$m\to 0$ (see \cite{pso22}). The Noether current itself makes sense as a
quantum field only if it is smeared with test functions $f_\mu$
satisfying
\bee
\int d^2x f_\mu(x)=0
\ee
On the other hand, it is not hard to see that $\phi_c(x)=curl(j)$ 
can be written as a function of the gradients:
\bee
\phi_c(x)=2\bigl((\partial_2\Phi^{(1)}(x))(\partial_1\Phi^{(2)}(x))-
(\partial_1\Phi^{(1)}(x))(\partial_2\Phi^{(2)}(x))\bigr)
\ee
and its two-point function is of the
form
\bee
\langle \phi_c(0)\phi_c(x)\rangle \propto {1\over (x^2)^2}
\ee

In other words, in this model we have found exactly the nontrivial
local cohomology class described in the previous section.
The problem in the massless contiuum limits of classical spin models
is then the following: it is conceivable that both {\it curl j} and 
{\it div j} have bona fide continuum limits, but the current itself does 
not. In other words, it could happen that there is a nontrivial second 
`local cohomology class' just as in the example discussed above.
But it turns out that reflection positivity can be used to rule out
such a possibility, provided we are dealing with a model that becomes
critical at a finite value of the inverse temperature $\beta$ (this is,
however, a prerequisite for constructing a massless continuum limit 
anyway). Our arguments show that both {\it curl j} and {\it div j} have 
correlations that are pure contact terms in the continuum limit; this 
means that in Minkowski space both the current and its dual are conserved,
thereby justifying Affleck's claim.

For completeness, let us mention an even more exotic possible way in which
the conformal classification of the critical behavior of the classical 
spin models could fail: one
could be imagine that the current itself has correlations that are pure
contact terms in the continuum, which would mean that the Noether
current simply vanishes as a quantum field. Of course this would also
imply vanishing of the corresponding charge, and since the commutator
of the charge with the (renormalized) spin field should be a component
of the (renormalized) spin field, those fields themselves would have to 
vanish, leading to a totally trivial theory containing only the vacuum.
There is a huge body of numerical results that makes this inconceivable,
and we also did some numerical simulations to eliminate this possibility
directly in the case of the $O(2)$ model \cite{pso21, pso22}.


\vskip4mm \noindent
{\bf 4. The Noether Current: Some Generalities}
\vskip2mm

The $O(N)$ model is determined by its standard Hamiltonian (action)

\bee
H=-\sum_{\langle ij\rangle} s(i)\cdot s(j)
\ee
where the sum is over nearest neighbor pairs on a square lattice
and the spins $s(.)$ are unit vectors in $\R^N$.
As usual Gibbs states are defined by using the Boltzmann factor
$\exp(-\beta H)$ together with the standard a priori measure on the
spins first in a finite volume, and then taking the thermodynamic
limit.

It is rigorously known \cite{FS} that for $N=2$ the model has a transition
to a massless spin wave phase at a certain $\beta=\beta_{KT}\approx 1.12$,
the so-called Kosterlitz-Thouless transition \cite{KT}. This transition
separates a high temperature phase with exponential clustering
from a low temperature one with only algebraic decay of correlations.
For $N>2$ the standard wisdom is that there is no such transition and
the model does not become critical at any finite $\beta$, but is
asymptotically free. For many years, however, we have been criticizing 
the arguments on which this standard wisdom is based and gave arguments
for an alternative scenario according to which ALL the $O(N)$ models have
a transition to a spin wave phase \cite{ps}.

Here we do not want to enter into this discussion, but we will produce
a criterion that distinguishes between these two scenarios.

But at first let us assume that our model has a finite critical point
and study the consequences. We are in particular interested in the 
correlations of the Noether currents, given by
\bee
j^{ab}_\mu(i)=\beta\Bigl(s^a(i)s^b(i+\hat\mu)
-s^a(i)s^b(i+\hat\mu)\Bigr)
\ee
Typically we will restrict ourselves to the case $a=1, b=2$ and omit
the flavor indices on the current.

On a torus the current can be decomposed into 3 pieces, a longitudinal,
a transverse and a constant (harmonic) piece. This
decomposition is  easiest in momentum space, and effected by the 
projections
\bee
P^T_{\mu\nu}=\Biggl(\delta_{\mu\nu}-{(e^{ip_\mu}-1)(e^{-ip_\nu}-1)\over
\sum_\alpha(2-2\cos p_\alpha)}\Biggr)(1-\delta_{p0}) ,
\ee

\bee
P^L_{\mu\nu=}={(e^{ip_\mu}-1)(e^{-ip_\nu}-1)\over
\sum_\alpha(2-2\cos p_\alpha)}(1-\delta_{p0})
\ee
and
\bee
P^h_{\mu\nu}=\delta_{\mu\nu}\delta_{p0}.
\ee
with $p_\mu=2\pi n_\mu/L$, $n_\mu=0,1,2,...,L-1$.

In the following we will mostly discuss these correlations in momentum
space. In particular we study the tranverse momentum space 2-point 
function
\bee
\hat F^T(p,L)\equiv \hat G(0,p;L)=\langle |\hat j_1(0,p)|^2\rangle
\ee
(for $p\neq 0$; the hat denotes the Fourier transform) \newline
and the longitudinal two-point function
\bee
\hat F^L(p,L)\equiv \hat G(p,0;L)=\langle |\hat j_1(p,0)|^2\rangle
\ee
(for $p\neq 0$).

Because the current is conserved, its divergence in the Euclidean world
should be a pure contact term, and for dimensional reasons the two-point 
function should be proportional to a $\delta$ function, i.e.
\bee
\hat F^L(p,L)=const.
\ee

The constant is in fact determined by a Ward identity in terms of
$E=\langle s(0)\cdot s(\hat\mu)\rangle$:
consider (for a suitable finite volume) the partition function
\bee
Z=\int\prod_i ds(i)\prod_{\langle ij\rangle}\exp\bigl(\beta s(i)\cdot
s(j)\bigr)
\ee
where $ds$ denotes the standard invariant measure on the $(N-1)$-sphere.
Replacing under the integral $s(i)$ by $\exp(\alpha L_{12})$, where
$L_{12}$ is an infinitesimal rotation in the 12 plane, does not change 
the integral. So expanding in powers of $\alpha$, all terms except the 
one of order $\alpha^0$ vanish indentically in $\alpha(i)$. This leads in
a well-known fashion to Ward identities expressing the conservation of 
the current. Looking specifically at the second order term in $\alpha$ 
and Fourier transforming, we obtain for all $p\neq 0$
\bee
\langle |j_1(p,0)|^2\rangle=\hat F^L(p,L)={2\over N} \beta E
\ee
This is confirmed impressively by the Monte Carlo simulations 
\cite{pso22}.

The thermodynamic limit is obtained by sending $L\to\infty$ for fixed 
$p=2\pi n/L$, so that in the limit $p$ becomes a continuous variable
ranging over the interval $[-\pi,\pi)$. The $O(N)$ models do
not show spontaneous symmetry breaking according to the Mermin-Wagner 
theorem, and presumably have a unique infinite volume limit at any
temperature.

The massive continuum limit is contructed as follows: First one takes the
thermodynamic limit of the the model in its high temperature phase. There
is a dynamically generated length scale $\xi$, the correlation length
regulating the exponential decay of the correlations. This is now taken
as the standard of length, and the fields are rescaled accordingly. In
particular the Noether current is rescaled as follows:
\bee
j_\mu^{ren}(x)=\xi j_\mu(i)
\ee
with $x=i/\xi$.
After that, the system is driven to the critical point, where 
$\xi\to\infty$.

The massless continuum limit, on the other hand, is obtained as follows: 
we take the thermodynamic limit of the model right at its critical point.
Since there is no dynamically generated scale, we take an arbitrary
sequence $l_n$ going to infinity as our standard of length. The currents
are then rescaled as
\bee
j_\mu^{ren}(x)=l_nj_\mu(i)
\ee
with $x=i/l_n$ and we take the limit $n\to\infty$.

\vskip4mm \noindent
{\bf 5. The Noether Current: Bounds and Inequalities}
\vskip2mm
The Gibbs measure formed with the standard action on the periodic lattice
has the property of reflection positivity (see for instance \cite{OSe}).
Reflection positivity means that expectation values of the form
\bee
\langle A\theta(A)\rangle,
\ee
are nonnegative, where $A$ is an observable depending on the spins
in the `upper half' of the lattice ($\{x|x_1>0\}$, and $\theta(A)$
is the complex conjugate of the same function of the spins at the sites 
with $x_1$ replaced by $-x_1$.
Applied to the current two-point functions this yields:
\bee
F^L(x_1,L)=\sum_{x_2} \langle j_1(x_1,x_2) j_1(0,0)\rangle\leq 0
\ee
for $x_1\neq 0$ and
\bee
F^T(x_1,L)=\sum_{x_2} \langle j_2(x_1,x_2) j_2(0,0)\rangle\geq 0
\ee
for all $x_1$.
From these two equations and eq.(27) it follows directly that
\bee
0\leq \hat F^T(p,L)\leq \hat F^T(0,L)=\hat F^L(0,L)\leq \hat F^L(p,L)
={2\over N}\beta E
\ee

These inequalities remain of course true in the thermodynamic limit, 
but we have to be careful with the order of the limits. If we define
$\hat F^T(p,\infty)$ and $\hat F^L(p,\infty)$ as the Fourier transforms
of $\lim_{L\to\infty} F^T(x,L)$ and $\lim_{L\to\infty} F^L(x,L)$,
respectively, one conclusion can be drawn immediately:

{\bf Proposition:}
$\hat F^T(p,\infty)$ and $\hat F^L(p,\infty)$ are continuous functions 
of $p\in [-\pi,\pi)$.

The proof is straightforward, because due to the inequalities (32)
(33) and (34) together with the finiteness of $\beta_{crt}$ the limiting
functions $F^L$ and $F^T$ in $x$-space are absolutely summable.
But it is not assured that the limits $L\to\infty$ and $p\to 0$ can 
be interchanged, nor that the thermodynamic limit and Fourier
transformation can be interchanged. On the contrary, by the numerics 
presented in \cite{pso21,pso22}, as well as finite size scaling arguments,
it is suggested that
\bee
\lim_{p\to 0}\lim_{L\to\infty} \hat F^L(p,L)>
\lim_{L\to\infty}\hat F^L(0,L)
\ee
and therefore also
\bee
\lim_{p\to 0}\lim_{L\to\infty} \hat F^L(p,L)
>\lim_{p\to 0}\lim_{L\to\infty} \hat F^T(p,L).
\ee

The fact that these are strict inequalities plays an important role in 
the justification of Affleck's claim, as will be seen below.

To continue, let us describe how the two types of continuum limit are 
taken in Fourier space, concretely for our functions $\hat F^T(p,\infty),
\hat F^L(p,\infty)$.

The massive continuum limit means considering $\hat F^T(p)$ etc. for
a sequence of $\xi$ values diverging to $\infty$ as functions of 
$q\equiv p/m=p\xi$, i.e. taking
\bee
\lim_{\xi\to\infty} \hat T(q)\equiv \hat F^T(qm)
\ee
In this context it is important to note that the functions $\hat F^T(p)$
depend explicitly on $\beta$ which is sent to $\beta_{crt}$, and through
this on the correlation length $\xi$, which is sent to $\infty$.

The massless continuum limit on the other hand is obtained by going to 
the critical point and considering $\hat F^T(p)$ etc. as a function of 
$q\equiv p/l_n$, i.e. taking
\bee
\lim_{n\to\infty} \hat T(q)\equiv \hat F^T(q/l_n)
\ee
In this case we are always dealing with only one function $\hat F^T(p)$,
not depending on $n$, because $\beta$ is fixed to its critical value.

\vskip4mm \noindent
{\bf 6. Consequences}
\vskip2mm

For the massless continuum limit the inequalities (33) lead to an
important consequence, which closes the main gap in Affleck's argument
by showing the triviality of the second local cohomology class defined
by the {\it curl} of the noether current:

{\bf Proposition:}
In the massless continuum limit both $\hat F^L(p,\infty)$ and 
$\hat F^T(p,\infty)$ converge to constants for $p\neq~0$.

{\bf Corollary:}
The local cohomology class defined by $curl(j)$ is trivial.

{\it Proof:}
Let $\hat F(p)$ be the Fourier transform of either $\hat F^T(p,\infty)$
or $\hat F^L(p,\infty)$. We consider $\hat F(p)$ as a distribution on
$[-\pi,\pi)$. We extend $\hat F(p)$ to a periodic distribution on the 
whole real line. The continuuum limit of $F(n)$ (the corresponding
function in $x$ space) also has to be considered in the sense of 
distributions. If we change our standard of length to $l_M=M$, the 
lattice spacing will be $a=1/M$, respectively.
For an arbitrary test function $f$ (infinitely differentiable and of
compact support) on the real axis we then have to consider the limit
$M\to\infty$  of
\bee
(F,f)_M\equiv \sum_n f(n/M) F(n).
\ee
We claim that the right hand side of this is equal to
\bee
{1\over 2\pi}\int_\infty^\infty dq \hat F({q/M})\hat f(q).
\ee
{\it Proof:} Insert in eq.(38)
\bee
F(n)={1\over 2\pi} \int_{-\pi}^\pi dp\hat F(p) e^{ipn}
\ee
and
\bee
f({n/M})={1\over 2\pi} \int_{-\pi}^\pi dp\hat f(p) e^{ipn/M}
\ee
and use the identity
\bee
\sum_n e^{ipn+iqnb}=2\pi\sum_r\delta(p+qb+2\pi r)
\ee
This produces, after carrying out the trivial integral over $q$ using
the $\delta$ distribution,
\bee
{M\over 2\pi} \int_{-\pi}^\pi dp \sum_{r=-\infty}^\infty
\hat F(-p)\hat f((p+2\pi r)M)
={1\over 2\pi}\sum_{r=-\infty}^\infty
\int_{-M\pi}^{M\pi} dq \hat F(-{q/M}) \hat f(q+2\pi Mr)
\ee
Finally, using the periodic extension of $\hat F(p)$, this becomes
what is claimed in eq.(39).

From eq.(39) one sees that what is relevant for the continuum limit is
the small momentum behavior of $\hat F(p)$. In particular, if
$\lim_{p\to 0}\hat F(p)\equiv \hat F(0)$ exists, we obtain
\bee
\lim_{M\to\infty} (F,f)_M={1\over 2\pi} \hat F(0) \int dq \hat f(q)
={1\over 2\pi}f(0)\hat F(0)
\ee
expressing the fact that in this case the limit of $F$ is a pure contact
term. This finishes the proof of the proposition.

In spite of this result, Affleck's claim could still fail in a different
way if $\hat F^T(p,\infty)$ and $\hat F^L(p,\infty)$ converged to the
same constant in the continuum limit. Let us denote the continuum limit
of $\hat F^T(p,\infty)$ by $g$. Then the current-current correlation
in this limit is
\bee
\langle j_\mu j_\nu\rangle\hat(p) = \beta E P^L_{\mu\nu}+
g P^T_{\mu\nu} = g\delta_{\mu\nu}+(\beta E-g){p_\mu p_\nu\over p^2}.
\ee
So we see that if $g=\beta E$, the current-current correlation reduces
to a pure contact term and vanishes in Minkowski space. Above we proved
only that
\bee
g\leq \beta E
\ee
But if the the current-current correlation were a pure contact term,
it would be unavoidable to conclude that also the spin field becomes
ultralocal. This can be seen as follows: if the current is ultralocal in 
the euclidean world, by the Osterwalder-Schrader reconstruction  \cite{OS}
the current field operator in Minkowski space has to vanish, and so does
the charge operator $Q_{12}=\int dx j_o(x,t)$. But if the charge operator
generates a global $O(N)$ symmetry, it has to have the following 
commutation relation with the (renormalized, Minkowskian) spin field
$s(x)$:
\bee
[Q_{12},s_a(x)]=0,\ \ a>2
\ee
\bee
[Q_{12},s_1(x)]=s_2(x)
\ee
\bee
[Q_{12},s_2(x)]=-s_1(x)
\ee
which would then imply that $s(x)$ vanishes identically. This argument
is not fully rigorous, because it assumes eq.(47) as well as the validity
of the Osterwalder-Schrader axioms; both have not been proven in full 
rigor for the continuum limit of the $O(N)$ models. Also there is only
numerical evidence, but no rigorous proof, that the continuum limit
of the spin field is not ultralocal.
For these reasons we presented in \cite{pso22}
numerical data which  (together with finite size scaling arguments) 
rule out directly ultralocality of the current.

Let us now turn to the massive continuum limit. For this the inequalities
(33) yield the announced bound on the transition temperature in terms
of the tranverse Noether current in momentum space:

{\bf Proposition:} For the $O(N)$ models the critical inverse temperature
satisfies 
\bee
\beta_{crt}\geq {N\over 2} \sup_p  \hat F^T(p)
\ee
The quantity $J(p)=\hat F^T(p)-\hat F^T(0)$ satisfies
\bee
J(p)\leq {2\over N} \beta_{crt}
\ee

{\it Proof:}
Both statements follow directly by taking first the thermodynamic and
then the massive continuum limit of eq.(33), using also the trivial
fact $E\leq 1$.

This result is the announced criterion distinguishing between 
$\beta_{crt}<\infty$ and $\beta_{crt}=\infty$ by the boundedness or
unboundedness of $J(p)$ or $\hat F^T(p)$. Of course it is a highly
nontrivial matter to verify this criterion. Balog and Niedermaier
\cite{bn} gave arguments that $J(p)$ is unbounded in their form factor
construction of the $O(3)$ model, which seems to suggest 
$\beta_{crt}=\infty$. But we found by very precise numerical simulations 
evidence \cite{ffmc} that the form factor construction disagrees with the (massive)
continuum limit of the lattice $O(3)$ model, leaving open the possibility
that indeed $\beta_{crt}<\infty$ as long advocated by us.

\vskip4mm \noindent

A.P is grateful to the Max-Planck-Institut for its hospitality
and financial support;
E.S. wishes to thank the University of Arizona for its hospitality
and financial support.

\end{document}